**Prime-Field PINI: Machine-Checked Composition Theorems for Post-Quantum NTT Masking**

**Ray Iskander**[1], **Khaled Kirah**[2,*]

[1] Verdict Security, ray@verdictsecurity.com
[2] Faculty of Engineering, Ain Shams University, Cairo, Egypt

**Abstract**

This is Paper 6 of a series of formally-verified analyses of masked NTT hardware for post-quantum cryptography; Paper 1 [1] established structural dependency analysis of the QANARY platform, and Paper 2 [2] quantified security margins under partial NTT masking. Boolean masking composition is well-understood through NI, SNI, and PINI. Arithmetic masking over $\mathbb{Z}_q$ for prime q, the foundation of NTT-based post-quantum cryptography, has lacked an analogous theory. We prove, to our knowledge, the first machine-checked composition theorems for arithmetic masking over prime fields. Our key insight is the renewal argument: when a fresh random mask is applied between two pipeline stages, the intermediate wire becomes perfectly uniform regardless of Stage 1's security parameter. For two PF-PINI gadgets with parameters $k_1$ and $k_2$, the composed two-stage pipeline with fresh masking satisfies PF-PINI( $k_2$ ), Stage 1's multiplicity is completely erased from the composed output. Without fresh masking, intermediate wires have multiplicity up to $k_1$, creating a necessary condition for differential power analysis. We formalize both theorems in Lean 4 with 18 machine-checked proofs and zero sorry stubs. We formally bridge the algebraic and hardware-faithful arithmetic models of Barrett reduction, and instantiate the theorems to formally diagnose Microsoft's Adams Bridge PQC accelerator: its absence of fresh inter-stage masking leaves Barrett output wires non-uniform under the first-order probing model, the same architectural flaw that two independent empirical analyses [3, 4] and our own prior structural analysis [1] identified. Computational evidence further suggests the 1-Bit Barrier is universal across Barrett and Montgomery reductions.

# 1. Introduction

The Number Theoretic Transform (NTT) is the computational backbone of the new NIST post-quantum standards ML-DSA (FIPS 204) and ML-KEM (FIPS 203). Hardware implementations of NTT-based cryptography must resist side-channel attacks, particularly differential power analysis (DPA), through masking: splitting sensitive values into random shares that are processed independently.

For Boolean masking over $\mathrm{GF}(2^n)$, the composition problem is solved. Ishai, Sahai, and Wagner established the foundational probing security model [5]. Barthe et al. formalized Non-Interference (NI) and Strong Non-Interference (SNI), proving that SNI gadgets compose securely under sequential composition [6]. Cassiers and Standaert introduced Probe Isolating

*Correspondence Author: khaled.kirah@eng.asu.edu.eg
Ray Iskander: ray@verdictsecurity.com

Non-Interference (PINI), achieving trivially composable gadgets, any PINI circuit built from PINI gadgets is automatically PINI [7]. These results have been formalized in EasyCrypt, providing machine-checked guarantees for Boolean masking [8].

### 1.1. The Arithmetic Gap

Arithmetic masking over $\mathbb{Z}_q$ for prime $q$ is fundamentally different. In $\mathrm{GF}(2^n)$, XOR is its own inverse, and masking $x = \tilde{x} \oplus m$ preserves the algebraic structure cleanly. In $\mathbb{Z}_q$, masking $x = \tilde{x} + m \pmod{q}$ interacts with modular reduction in non-trivial ways. Barrett reduction, for example, has input-dependent branching: the map $x \mapsto ((x + 2^s - m) \bmod 2^s) \bmod q$ has a two-branch structure that breaks uniform mask distributions. This is the PF-PINI(2) phenomenon, some output values are hit by two masks while others are hit by one, leaking up to one bit of information per probed wire. No prior work provides composition theorems for arithmetic masking over prime fields, and none provides machine-checked proofs in this setting. This paper fills both gaps.

### 1.2. The Renewal Argument

Our key insight is remarkably simple: a single fresh random mask applied between pipeline stages completely erases Stage 1's security parameter from the composed output.

Consider a two-stage pipeline $\mathrm{G}_1 \rightarrow \mathrm{G}_2$ where a fresh mask $\mathrm{m}_{\mathrm{fresh}}$ is applied between stages. For any target intermediate value w, the set of mask pairs $(\mathrm{m}_1, \mathrm{m}_{\mathrm{fresh}})$ that produce w has cardinality exactly q. The proof is four lines of Lean 4: for each $\mathrm{m}_1$, there is exactly one $\mathrm{m}_{\mathrm{fresh}} = \mathrm{G}_1(\mathrm{x}, \mathrm{m}_1) - \mathrm{w}$ that works, and the map $\mathrm{m}_1 \mapsto (\mathrm{m}_1, \mathrm{G}_1(\mathrm{x}, \mathrm{m}_1) - \mathrm{w})$ is injective.

This renewal lemma has a powerful consequence: Stage 2 sees each input value with equal frequency, regardless of how non-uniform Stage 1's output is. The composed pipeline's PF-PINI parameter is therefore bounded by $\mathrm{k}_2$ alone, $\mathrm{k}_1$ is completely erased.

### 1.3. Contributions

We make six contributions:

1. **The Renewal Theorem** (Theorem 4.1): Fresh inter-stage masking erases Stage 1's PF-PINI parameter from the intermediate wire distribution. Every intermediate value is produced by exactly $q$ mask pairs.
2. **The Positive Composition Theorem** (Theorem 4.4): The composed two-stage pipeline with fresh masking satisfies PF-PINI($k_2$). The composed output multiplicity is bounded by $k_2 \cdot q^2$ over the three-mask space $\mathbb{Z}_q^3$.
3. **The Negative Theorem** (Theorems 5.1–5.3): Without fresh masking, intermediate wires have multiplicity up to $k_1$, and when this bound is achieved, the wire is non-uniform, creating a necessary condition for DPA.
4. **The Algebraic-Hardware Bridge** (Theorem 6.1): The algebraic Barrett reduction map equals the hardware-faithful Nat arithmetic map under the scope condition $q \leq 2^s$, so the PF-PINI(2) bound applies directly to hardware implementations.
5. **Adams Bridge Diagnosis** (Section 7): We formally instantiate both composition theorems to diagnose the architectural flaw in Microsoft's Adams Bridge PQC accelerator, the same flaw that three independent empirical analyses exploited [1, 3, 4].
6. **Machine-Checked Proofs**: All results are formalized in Lean 4 with zero sorry stubs: a total of 18 public theorems (3 combinatorial lemmas, 4 positive-composition results, 4 negative-composition results, 5 Adams Bridge instantiations, 2 algebraic-hardware bridge results). The proof artifact is publicly available.

### 1.4. Paper Organization

Section 2 reviews related work on masking composition. Section 3 defines PF-PINI and establishes its relationship to standard notions. Section 4 proves the renewal theorem and positive composition. Section 5 proves the negative results. Section 6 bridges algebraic and hardware Barrett models. Section 7 instantiates the theorems for Adams Bridge. Section 8 discusses extensions and limitations. Section 9 concludes.

## 2. Related Work

### 2.1. Boolean Masking Composition

The composition theory for Boolean masking is mature. Reference [5] established the probing security model, proving a construction that resists $t$-probing attacks. Reference [6] formalized this as Non-Interference (NI) and Strong Non-Interference (SNI), proving that SNI gadgets compose securely: the output of an SNI gadget can be fed to another gadget without degrading the security order. In reference [7] PINI (Probe Isolating Non-Interference) was introduced, achieving trivially composable gadgets. Any circuit built from PINI components is automatically PINI, with no composition proof required. These results provide a complete theoretical foundation for Boolean masking composition. However, they operate over $\mathrm{GF}(2^n)$ and do not extend to arithmetic masking over $\mathbb{Z}_q$ for prime $q$.

### 2.2. Automated Verification Tools

Several tools automate the verification of masked implementations. maskVerif [9] verifies higher-order masking in the presence of physical defaults such as glitches and transitions. SILVER [10] uses Reduced Ordered Binary Decision Diagrams (ROBDDs) for statistical independence verification at the gate level. Coco [11] co-verifies masked software on CPUs down to the gate level. These tools verify individual gadgets or circuits against the probing model, but they do not provide composition theorems, general results about how the security of composed pipelines relates to the security of individual stages.

### 2.3. Formal Verification of Masking

Machine-checked proofs of masking security have been developed in EasyCrypt [8], providing high-assurance guarantees for Boolean masking constructions. These formalizations cover NI, SNI, and specific masking schemes, but they operate exclusively over $\mathrm{GF}(2^n)$. No EasyCrypt (or other proof assistant) formalization addresses arithmetic masking composition over $\mathbb{Z}_q$.

### 2.4. Arithmetic Masking

References [12] developed secure conversion between Boolean and arithmetic masking, and [13] achieved conversion with logarithmic complexity. These works address the construction and conversion of arithmetic masking but do not provide composition theorems, they do not analyze what happens when arithmetic gadgets are chained in pipelines.

### 2.5. PQC Hardware and Empirical Attacks

The Adams Bridge PQC accelerator [14] implements masked NTT for both ML-DSA and ML-KEM using Domain-Oriented Masking (DOM). Reference [3] demonstrated DPA on Adams Bridge's masked BFU multiplier using 10,000 power traces, confirming the multiplicity-2 phenomenon. In reference [4] systematic masking flaws through code review was identified. The authors in [1] identified 14 physically exploitable vulnerability instances across 5 modules using structural dependency analysis. Our work provides the formal theoretical explanation for why these attacks succeed: the absence of fresh inter-stage masking.

### 2.6. The Gap We Fill

No prior work proves composition theorems for arithmetic masking over prime fields. No prior work provides machine-checked proofs of masking composition in any proof assistant other than EasyCrypt, and EasyCrypt's results are limited to Boolean masking over $GF(2^n)$. We fill both gaps: general composition theorems for $\mathbb{Z}_q$, formalized in Lean 4 with zero sorry stubs.

## 3. Preliminaries

### 3.1. Arithmetic Masking

Let q be a prime. A value $x \in \mathbb{Z}_q$ is arithmetically masked by a random mask $m \in \mathbb{Z}_q$ as $\tilde{x} = x - m \pmod{q}$. A masked gadget G takes a secret x and a mask m and produces an output $G(x, m) \in \mathbb{Z}_q$. The security of the gadget depends on how the output distribution $\{G(x, m): m \in \mathbb{Z}_q\}$ varies with the secret x.

**Scope note.** Although our Lean formalization requires only $q \geq 1$ (the structure $(\mathbb{Z}_q, +, -)$ as a finite abelian group), we focus on prime q throughout this paper because that is the setting of the NIST PQC standards ML-KEM ($q = 3329$) and ML-DSA ($q = 8{,}380{,}417$). The composition theorems of Sections 4–5 generalize to any finite abelian group with subtraction.

### 3.2. The PF-PINI(k) Definition

We define the Prime-Field PINI parameter as a quantitative measure of single-wire security.

**Definition 3.1** (PF-PINI Gadget). An PF-PINI gadget over $\mathbb{Z}_q$ is a triple (compute, k,bound) where:

- compute: $\mathbb{Z}_q \times \mathbb{Z}_q \to \mathbb{Z}_q$ is the gadget's computation,
- $k \in \mathbb{N}$ is the PF-PINI parameter (called maxMult in our Lean formalization),
- bound: $\forall x, v \in \mathbb{Z}_q,\ |\{m \in \mathbb{Z}_q : \text{compute}(x, m) = v\}| \leq k$.

In Lean 4, this is the PFPINIGadget structure:

```
structure PFPINIGadget (q : ℕ) [NeZero q] where
  compute : ZMod q → ZMod q → ZMod q
  maxMult : ℕ
  bound : ∀ x v, (univ.filter (fun m => compute x m = v)).card ≤ maxMult
```

### 3.3. Normalization and Interpretation

For a gadget with mask space $\mathbb{Z}_q^n$ of size $q^n$, the uniform baseline is $q^{n-1}$: if the output were uniformly distributed, each output value would be hit by exactly $q^{n-1}$ mask tuples. PF-PINI(k) means the maximum multiplicity is at most $k \cdot q^{n-1}$. The parameter k measures the multiplicative deviation from uniformity:

- $k = 1$: perfectly uniform output distribution.
- $k = 2$: some outputs hit by up to $2 \times$ the uniform rate (up to 1 bit of leakage per wire).
- $k > 1$: at most $\log_2(k)$ bits of information per probed wire.

### 3.4. Relationship to Standard Notions

PF-PINI connects to the standard masking security hierarchy as follows.

**Single-wire vs. multi-probe.** PF-PINI(k) is a single-wire metric under the first-order probing model. It bounds the distribution of one output wire over masks. This is the single-wire analogue of 1-NI: PF-PINI(1) implies that each individual wire independently satisfies 1-probing security, the output distribution is uniform and independent of the secret x for that wire. Multi-probe security, where an attacker probes multiple wires simultaneously, is the natural extension and is left to future work (Section 8).

**Quantitative vs. binary.** NI, SNI, and PINI are binary notions: a gadget either satisfies the property or does not. PF-PINI is quantitative: it measures how much a wire leaks (at most $\log_2(k)$ bits), not merely whether it leaks. This quantitative information is essential for composition: the positive composition theorem (Theorem 4.3) shows that the composed pipeline's parameter depends only on $k_2$, regardless of $k_1$.

**Distinction from PINI.** Despite the name, PF-PINI is fundamentally different from PINI [7]. PINI provides multi-probe guarantees by construction: any PINI circuit built from PINI gadgets is PINI. PF-PINI provides single-wire multiplicity bounds with explicit composition theorems (this paper). PF-PINI also does not distinguish between input-dependent and output-dependent probes, unlike SNI [6].

**Comparison to statistical distance.** Cassiers and Standaert's work [7] uses statistical distance from the uniform distribution as a leakage metric. PF-PINI uses maximum multiplicative deviation, which has the advantage of composing cleanly through the fiber decomposition argument (Section 4).

### 3.5. Pipeline Model

We study two-stage pipelines $G_1 \to G_2$ in two configurations.

**With fresh masking.** The composed computation is:

$$\text{composedWithFresh}(G_1, G_2, x, m_1, m_{\text{fresh}}, m_2) = G_2(G_1(x, m_1) - m_{\text{fresh}}, m_2)$$

**Without fresh masking.** The composed computation is:

$$\text{composedNoFresh}(G_1, G_2, x, m_1, m_2) = G_2(G_1(x, m_1), m_2)$$

All masks $m_1, m_{\text{fresh}}, m_2$ are drawn independently and uniformly from $\mathbb{Z}_q$.

### 3.6. Known PF-PINI Results

Our composition theorems build on two prior machine-checked results:

- **NTT Butterfly**: The identity gadget $\text{compute}(x, m) = x - m$ satisfies PF-PINI(1). This models the butterfly operation in masked NTT, where each output wire is a bijective function of the mask.
- **Barrett Reduction**: The Barrett internal map satisfies PF-PINI(2). The two-branch structure of modular reduction, $x - m$ when $m \leq x$, and $x - m + 2^s$ when $m > x$, creates a maximum multiplicity of 2. This is the 1-Bit Barrier: a fundamental property of modular reduction, not a design choice.

## 4. The Renewal Theorem

This section presents the paper's central results: the renewal lemma and the positive composition theorem.

### 4.1. The Renewal Lemma

**Theorem 4.1** (Renewal Lemma; fresh_mask_renewal). Let $G_1$ be any PF-PINI gadget over $\mathbb{Z}_q$, and let $x, w \in \mathbb{Z}_q$. Then:

$$|\{(m_1, m_{fresh}) \in \mathbb{Z}_q^2 : G_1(x, m_1) - m_{fresh} = w\}| = q$$

Proof. For each $m_1 \in \mathbb{Z}_q$, setting $m_{fresh} = G_1(x, m_1) - w$ is the unique solution. The map $\varphi: m_1 \mapsto (m_1, G_1(x, m_1) - w)$ is injective (the first component determines the pair). The filter set equals $\text{Image}(\varphi)$, and:

$$|\text{Image}(\varphi)| = |\mathbb{Z}_q| = q$$

since $\varphi$ is injective and its domain is all of $\mathbb{Z}_q$. ▫

In Lean 4, the complete proof is:

```
theorem fresh_mask_renewal
   (G₁ : PFPINIGadget q) (x w : ZMod q) :
   (univ.filter (fun p : ZMod q × ZMod q =>
     G₁.compute x p.1 - p.2 = w)).card = card (ZMod q) := by
 have h_eq : ... = univ.image (fun m₁ => (m₁, G₁.compute x m₁ - w)) := by
   ext ⟨m₁, mf⟩; simp; constructor
   · intro h; exact ⟨m₁, rfl, by linear_combination h⟩
   · rintro ⟨_, rfl, hmf⟩; linear_combination hmf
 have h_inj : Function.Injective (fun m₁ => (m₁, G₁.compute x m₁ - w)) := by
   intro a b hab; exact (Prod.mk.inj hab).1
 rw [h_eq, card_image_of_injective _ h_inj, card_univ]
```

The renewal lemma is strikingly general. It makes no assumption about $G_1$'s structure, even a maximally non-uniform gadget with $k_1 = q$ produces perfectly uniform intermediate values after fresh masking. Moreover, the argument generalizes to masking over any finite group where masking is by group subtraction: the proof only requires that the group has q elements and subtraction is well-defined.

**Corollary 4.2** (Intermediate Uniformity; fresh_mask_uniform). The intermediate wire $G_1(x, m_1) - m_{fresh}$ is perfectly uniform: for any $w_1, w_2 \in \mathbb{Z}_q$,

$$|\{(m_1, m_{fresh}) : G_1(x, m_1) - m_{fresh} = w_1\}| = |\{(m_1, m_{fresh}) : G_1(x, m_1) - m_{fresh} = w_2\}|$$

Proof. Both sides equal q by Theorem 4.1. ▫

### 4.2. Fiber Decomposition

The composition theorems rely on a standard combinatorial lemma that we formalize as a reusable building block.

**Lemma 4.3** (Fiber Decomposition; card_filter_prod_le_mul). Let $\alpha, \beta$ be finite types and $P: \alpha \times \beta \to \text{Prop}$. If for every $a \in \alpha$, the fiber $|\{b \in \beta : P(a, b)\}| \leq k$, then:

$$|\{(a,b) \in \alpha \times \beta: P(a,b)\}| \leq |\alpha| \cdot k$$

Proof. Decompose by fibers: $\sum_{a\in\alpha} |\{b: P(a,b)\}| \leq \sum_{a\in\alpha} k = |\alpha| \cdot k$. ▫

### 4.3. The Positive Composition Theorem

**Theorem 4.4** (Positive Composition; pfpini_composition_with_fresh_mask). Let $G_1$ be PF-PINI($k_1$) and $G_2$ be PF-PINI($k_2$). With fresh inter-stage masking, the composed output multiplicity over the mask space $\mathbb{Z}_q^3$ satisfies:

$$|\{(m_1, m_{fresh}, m_2) \in \mathbb{Z}_q^3: G_2(G_1(x, m_1) - m_{fresh},\ m_2) = v\}| \leq k_2 \cdot q^2$$

The uniform baseline for $\mathbb{Z}_q^3$ mapping to $\mathbb{Z}_q$ is $q^2$. Therefore, the composed pipeline satisfies PF-PINI($k_2$).

Proof. Reassociate $\mathbb{Z}_q \times \mathbb{Z}_q \times \mathbb{Z}_q$ as $(\mathbb{Z}_q \times \mathbb{Z}_q) \times \mathbb{Z}_q$, treating $(m_1, m_{fresh})$ as the outer type and $m_2$ as the inner type. For each fixed $(m_1, m_{fresh})$, the input to $G_2$ is some value $w = G_1(x, m_1) - m_{fresh}$, and by $G_2$'s PF-PINI($k_2$) bound:

$$|\{m_2: G_2(w, m_2) = v\}| \leq k_2$$

Applying the fiber decomposition lemma (Lemma 4.3) with $\alpha = \mathbb{Z}_q^2$ and $\beta = \mathbb{Z}_q$:

$$|\{(m_1, m_{fresh}, m_2): G_2(G_1(x, m_1) - m_{fresh}, m_2) = v\}| \leq q^2 \cdot k_2 \quad ▫$$

**Key observation.** The renewal lemma (Theorem 4.1) is not needed for the output bound, the fiber decomposition alone suffices. The renewal lemma's role is proving intermediate wire uniformity (Corollary 4.2): a qualitative property distinct from the quantitative output bound. The positive composition theorem establishes that the composed output satisfies PF-PINI($k_2$) with no dependence on $k_1$.

**Corollary 4.5** (Symmetric Bound; pfpini_composition_max_bound).

$$|\{(m_1, m_{fresh}, m_2):\text{composedWithFresh}(G_1, G_2, x, m_1, m_{fresh}, m_2) = v\}| \leq \max(k_1, k_2) \cdot q^2$$

Proof. Since $k_2 \leq \max(k_1, k_2)$, the result follows from Theorem 4.4. ▫

The symmetric bound is a convenience corollary. The tight bound $k_2$ is the headline result: fresh masking erases Stage 1's contribution entirely.

## 5. Composition Without Fresh Masking

We now prove that without fresh masking, intermediate pipeline wires are exposed to probing attacks. The contrast between Sections 4 and 5 is the paper's punchline: the security difference lies entirely in the intermediate wire.

### 5.1. Intermediate Wire Exposure

**Theorem 5.1** (Intermediate Wire Multiplicity; intermediate_wire_multiplicity_bound). *Without fresh masking, the intermediate wire $G_1(x, m_1)$ has multiplicity bounded by $k_1$:*

$$|\{m_1 \in \mathbb{Z}_q: G_1(x, m_1) = v\}| \leq k_1$$

*Proof.* This is simply the PF-PINI($k_1$) bound applied to $G_1$. ▫

For Barrett reduction ($k_1 = 2$), some intermediate values are hit by 2 masks while others are hit by 1 or 0. An attacker probing this wire observes a non-uniform distribution that depends on the secret $x$, gaining up to 1 bit of information per observation.

### 5.2. Output Bound Without Fresh Masking

**Theorem 5.2** (No-Fresh Output Bound; composed_no_fresh_output_bound). Without fresh masking, the composed output multiplicity over $\mathbb{Z}_q^2$ satisfies:

$$|\{(m_1, m_2) \in \mathbb{Z}_q^2 : G_2(G_1(x, m_1),\ m_2) = v\}| \leq k_2 \cdot q$$

Proof: Apply the fiber decomposition (Lemma 4.3) with $\alpha = \mathbb{Z}_q$, $\beta = \mathbb{Z}_q$. For each fixed $m_1$, the value $w = G_1(x, m_1)$ is determined, and $|\{m_2 : G_2(w, m_2) = v\}| \leq k_2$. Therefore $|\{(m_1, m_2)$:composedNoFresh$(G_1, G_2, x, m_1, m_2) = v\}| \leq q \cdot k_2$. ▫

### 5.3. The Security Gap

**Theorem 5.3** (Security Gap; security_gap_intermediate). If $k_1 > 1$ and the PF-PINI bound is achieved (there exists v with $|\{m_1 : G_1(x, m_1) = v\}| = k_1$), then there exists an intermediate value with multiplicity greater than 1:

$$\exists\, v \in \mathbb{Z}_q : |\{m_1 : G_1(x, m_1) = v\}| > 1$$

Proof: Take v achieving the bound; since $k_1 > 1$, the count is $> 1$. ▫

> **Crucial Insight.** The output satisfies PF-PINI($k_2$) in both cases, with and without fresh masking. The security difference is entirely in the intermediate wire:

| | **With Fresh Mask** | **Without Fresh Mask** |
|---|---|---|
| **Output multiplicity** | $\leq k_2 \cdot q^2$ | $\leq k_2 \cdot q$ |
| **Output PF-PINI parameter** | $k_2$ | $k_2$ |
| **Intermediate wire** | **Uniform** (count $= q$) | **Non-uniform** (count $\leq k_1$) |
| **DPA on intermediate** | **Not possible** (uniform) | **Possible** ($\leq \log_2 k_1$ bits) |

Fresh masking does not improve output security, it protects the internal pipeline state. This is exactly why DPA succeeds on Adams Bridge: the attacker probes the intermediate wire between butterfly and Barrett stages.

## 6. Bridging Algebra and Hardware

The theorems in Sections 4–5 operate on algebraic maps over $\mathbb{Z}_q$. Hardware implementations compute in fixed-width natural number arithmetic with modular reduction. This section formally bridges the two models for Barrett reduction, ensuring that the PF-PINI(2) bound proved algebraically applies to the actual hardware computation.

### 6.1. The Verification Stack

The gap between formal proofs and silicon has three layers:

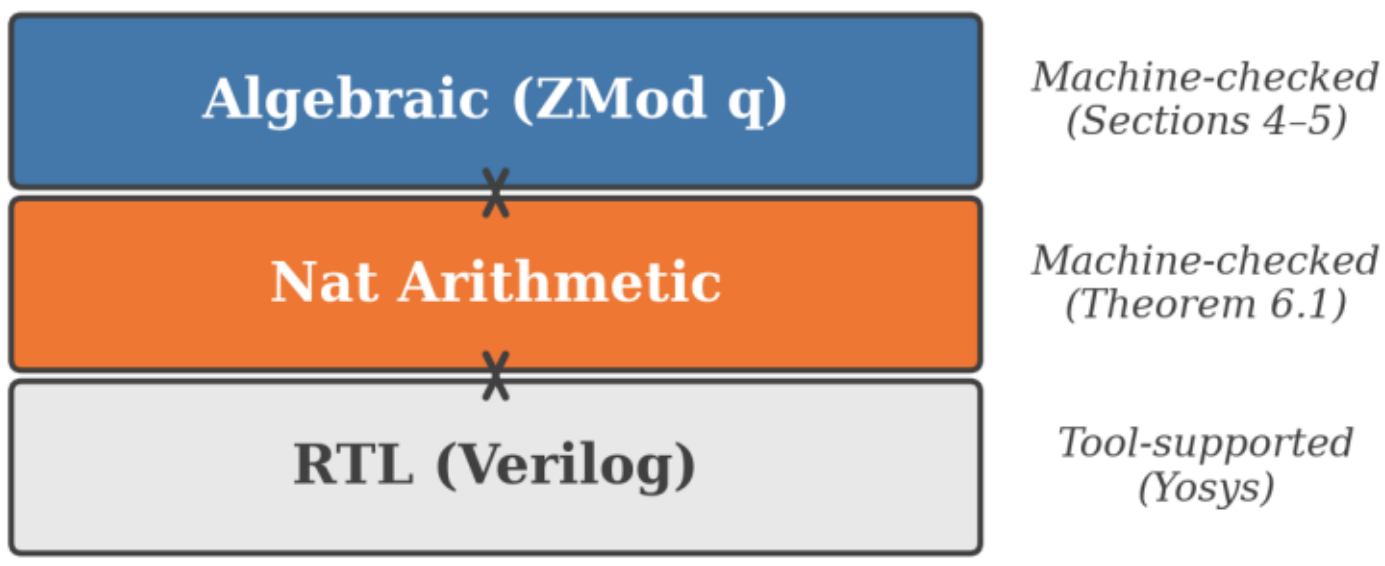

*Figure 1. The verification stack. Sections 4–5 prove composition theorems at the algebraic level (ZMod q). Theorem 6.1 bridges to hardware-faithful Nat arithmetic. The Nat-to-RTL step is supported by standard synthesis tools.*

Our formal bridge covers the mathematically subtle middle layer, the algebraic-to-Nat correspondence. The Nat-to-RTL step is well-understood and supported by standard synthesis tools (e.g., Yosys).

### 6.2. The NatEquivalence Theorem

Barrett reduction has two equivalent formulations. The *algebraic* map operates in $\mathbb{Z}_q$:

$$\text{barrettInternalMap}(s, x, m) = \begin{cases} x - m & \text{if } m.\text{val} \leq x.\text{val} \\ x - m + (2^s \bmod q) & \text{if } m.\text{val} > x.\text{val} \end{cases}$$

The *hardware-faithful* map operates in $\mathbb{N}$:

$$\text{barrettInternalMapNat}(s, x, m) = ((x.\text{val} + 2^s - m.\text{val}) \bmod 2^s) \bmod q$$

**Theorem 6.1** (NatEquivalence; barrett_nat_equivalence). *For all $x, m \in \mathbb{Z}_q$ and $s \in \mathbb{N}$ with $q \leq 2^s$:*

$$\text{barrettInternalMap}(s, x, m) = \text{barrettInternalMapNat}(s, q \leq 2^s, x, m)$$

*Proof.* Case split on $m.\text{val} \leq x.\text{val}$.

*Case 1 (no wraparound):* When $m.\text{val} \leq x.\text{val}$, the Nat computation satisfies $(x.\text{val} + 2^s - m.\text{val}) \bmod 2^s = x.\text{val} - m.\text{val}$ because $x.\text{val} - m.\text{val} < q \leq 2^s$. The subsequent $\bmod\ q$ is the identity since $x.\text{val} - m.\text{val} < q$. Both maps yield $x - m$ in $\mathbb{Z}_q$.

*Case 2 (wraparound):* When $m.\text{val} > x.\text{val}$, the sum $x.\text{val} + 2^s - m.\text{val}$ is positive (since $m.\text{val} < q \leq 2^s$) and less than $2^s$ (since $x.\text{val} < m.\text{val}$), so $\bmod\ 2^s$ is the identity. The result is $x.\text{val} + 2^s - m.\text{val}$, and $\bmod\ q$ reduces this modulo $q$, matching the algebraic $x - m + (2^s \bmod q)$ in $\mathbb{Z}_q$. ▫

**Corollary 6.2** (barrettNat_pfpini_two). *The hardware-faithful Barrett map satisfies PF-PINI(2):*

$$\forall\, x, v \in \mathbb{Z}_q \colon |\{m \colon \text{barrettInternalMapNat}(s, q \leq 2^s, x, m) = v\}| \leq 2$$

*Proof.* By Theorem 6.1, the filter sets are identical, so the bound transfers from the algebraic proof. ▫

---

## 7. Application: Adams Bridge Diagnosis

We instantiate both composition theorems to formally diagnose the Adams Bridge PQC accelerator [14], which implements masked NTT for ML-DSA (FIPS 204) and ML-KEM (FIPS 203).

### 7.1. The Pipeline

Adams Bridge chains two stages in its NTT pipeline: - **Stage 1 (Butterfly):** The NTT butterfly operation, modeled as the identity gadget with PF-PINI(1). The butterfly output wire is a bijective function of the mask, perfectly uniform. - **Stage 2 (Barrett):** Barrett modular reduction with PF-PINI(2). The two-branch structure creates a maximum multiplicity of 2.

The butterfly's PF-PINI(1) reflects good design at the individual gadget level. The Barrett PF-PINI(2) is the *1-Bit Barrier*, a fundamental property of modular reduction over $\mathbb{Z}_q$, not a design error. The issue is architectural: Adams Bridge applies *zero* fresh inter-stage masking between butterfly and Barrett stages.

### 7.2. Formal Diagnosis

We prove five theorems, each a direct instantiation of the general results.

**Theorem 7.1** (adams_bridge_butterfly_wire_ok). *The butterfly output wire has multiplicity* $\leq 1$*: perfectly uniform.*

**Theorem 7.2** (adams_bridge_barrett_wire_leaks). *The Barrett output wire has multiplicity* $\leq 2$*: up to 1 bit of leakage per probed wire under the first-order probing model.*

**Theorem 7.3** (adams_bridge_with_fresh_secure). *With fresh masking, the composed pipeline's output multiplicity is* $\leq \max(1,2) \cdot q^2 = 2q^2$.

**Theorem 7.4** (adams_bridge_fresh_pfpini_parameter). *The pipeline PF-PINI parameter with fresh masking is* $\max(1,2) = 2$.

**Theorem 7.5** (adams_bridge_intermediate_uniform_with_fresh). *With fresh masking, the intermediate wire between butterfly and Barrett is perfectly uniform: for all* $w \in \mathbb{Z}_q$*, the count of valid mask pairs equals* $q$.

### 7.3. Prescription

Adding one fresh mask per pipeline stage gives: - **Pipeline PF-PINI parameter:** 2 (same as Barrett alone) - **Intermediate wires:** Perfectly uniform (no DPA surface) - **Cost:** One additional $\mathbb{Z}_q$ random mask register and one subtraction per pipeline stage

> **Design Guideline.** For any masked NTT pipeline chaining PF-PINI($k_1$) and PF-PINI($k_2$) stages: insert one fresh random mask register between stages. Cost: one $\mathbb{Z}_q$ register + one subtraction per stage. Benefit: intermediate wire uniformity; composed pipeline satisfies PF-PINI($k_2$).

### 7.4. Convergent Evidence

Our formal diagnosis aligns with three independent empirical analyses that identified the same architectural flaw through different methods (Table 1):

| Analysis | Method | Year | Finding |
|---|---|---|---|
| Karabulut & Azarderakhsh [3] | 10,000 CPA traces | 2025 | DPA on BFU multiplier wire |
| Saarinen [4] | Code review | 2025 | Systematic masking flaws |
| Iskander & Kirah [1] | Structural analysis | 2025 | 14 physical vulnerability instances |
| **This paper** | **Machine-checked proof** | **2026** | **Barrett wires non-uniform; fresh mask fixes it** |

**Table 1.** Convergence of four independent methods identifying the same architectural flaw in Microsoft's Adams Bridge PQC accelerator.

The convergence of four complementary methods from three independent research groups, empirical power analysis [3], manual code review [4], automated structural analysis [1], and now machine-checked formal proof, provides strong evidence that the root cause is correctly identified.

## 8. Discussion and Extensions

### 8.1. Montgomery Reduction

Computational testing indicates that Montgomery reduction also satisfies PF-PINI(2). The map $(x + R - m)$ mod $R$ mod $q$ for $R = 2^s$ has the same two-branch structure as Barrett. We tested exhaustively for all primes $q \leq 31$ and for ML-KEM ($q = 3329$), and by sampling for ML-DSA ($q = 8{,}380{,}417$); see Table 2.

| Prime $q$ | $s$ | Max multiplicity | Verification |
|---|---|---|---|
| All $\leq 31$ | varies | 2 | Exhaustive |
| 3329 (ML-KEM) | 12 | 2 | Exhaustive |
| 8,380,417 (ML-DSA) | 23 | 2 | 100 random $x$, exhaustive $m$ |

**Table 2.** Computational evidence that Montgomery reduction satisfies PF-PINI(2) across primes used by NIST PQC standards. The full Lean proof is left to future work; see Section 4 for the analogous Barrett proof.

This suggests the 1-Bit Barrier is universal across standard modular reductions: any reduction with a mod $2^s$ mod q structure will have PF-PINI(2). Formal Lean proof of Montgomery PF-PINI(2) would follow the same argument as Barrett and is left to future work.

### 8.2. Multi-Stage Pipelines

Our composition theorems handle two-stage pipelines. Real NTT pipelines have $O(\log n)$ stages (7 for ML-KEM, 8 for ML-DSA). The inductive extension is conceptually straightforward: a composed gadget with fresh masking can be wrapped into a new PF-PINI gadget by defining an expanded mask type. However, the current PFPINIGadget type takes a single mask in $\mathbb{Z}_q$; the composed gadget takes three masks in $\mathbb{Z}_q^3$. A generalized PFPINIGadget parameterized by mask dimension would enable recursive composition.
The two-stage result is the hard part, the renewal argument is the key insight. For the Adams Bridge diagnosis, two-stage composition suffices (butterfly → Barrett is the critical pipeline).

### 8.3. Multi-Probe Extensions

PF-PINI bounds the multiplicity of a single wire. The natural extension to multi-probe security would bound joint distributions across multiple wires simultaneously, analogous to t-NI and t-SNI for Boolean masking. Specifically, a t-PF-PINI notion would bound:

$$|\{m: (G(x,m)|_{w_1}, \dots, G(x,m)|_{w_t}) = (v_1, \dots, v_t)\}|$$

for any set of $t$ probed wires. This is the natural next step in the theory.

### 8.4. Tightness

The bound PF-PINI($k_2$) for the composed pipeline is an upper bound. We do not prove a matching lower bound: there may exist gadget pairs where the actual maximum multiplicity is strictly less than $k_2 \cdot q^2$. Proving tightness, constructing a gadget pair that achieves the bound, is left to future work. Upper bounds without matching lower bounds are standard for first composition results in the masking literature [6, 7].

### 8.5. Implications for Standards

Machine-checked composition proofs could inform the FIPS 140-3 side-channel evaluation methodology. As post-quantum algorithms move into certified hardware, formal guarantees about pipeline composition provide a rigorous foundation for security evaluation beyond empirical testing.

### 8.6. The Six-Paper Program

This paper completes a formal theory of masked NTT hardware security:

1. **Paper 1** [1]: Structural dependency analysis identifies 14 vulnerability instances in Adams Bridge
2. **Paper 2** [2]: Security margin analysis with belief propagation attack simulation
3. **Paper 3** [15]: Ring axioms and value independence for $\mathbb{Z}_q$
4. **Paper 4** [16]: NTT butterfly satisfies PF-PINI(1)
5. **Paper 5** [17]: Barrett reduction satisfies PF-PINI(2); the 1-Bit Barrier
6. **This paper**: Composition theorems + Adams Bridge diagnosis + prescription

The loop is closed: vulnerability discovered, root cause formalized, fix proved.

## Conclusion

We prove, to our knowledge, the first machine-checked composition theorems for arithmetic masking over prime fields. Fresh inter-stage masking makes two-stage pipelines composable by erasing Stage 1's PF-PINI parameter from the composed output, its absence leaves intermediate wires non-uniform, creating conditions necessary for side-channel analysis. The renewal theorem, combined with per-gadget PF-PINI bounds, provides formally verified composition guarantees for single-wire security of masked NTT hardware pipelines.

## Code and Data Availability

The Lean 4 artifact accompanying this paper is publicly available under the MIT license:

- **Repository:** https://github.com/rayiskander2406/qanary-pf-pini-composition-arXiv-2604.25878
- **Toolchain:** Lean 4 v4.30.0-rc1 (managed via elan)
- **Pinned dependency:** Mathlib at commit 322515540d7fd29ef8992b82c89044f86f02ac10
- **License:** MIT (artifact); CC-BY-4.0 (manuscript)

### Reproduction

```
# Sibling repos required for local-path dependencies declared in lakefile.lean
mkdir qanary-artifacts && cd qanary-artifacts
git clone https://github.com/rayiskander2406/qanary-universal-masking-proofs-arXiv-2604.18717 qanary-universal
git clone https://github.com/rayiskander2406/qanary-masked-ntt-pipeline-security-arXiv-2604.20793 qanary-paper4
git clone https://github.com/rayiskander2406/qanary-one-bit-barrier-arXiv-2604.24670 qanary-paper5
git clone https://github.com/rayiskander2406/qanary-pf-pini-composition-arXiv-2604.25878 qanary-paper6

cd qanary-paper6
lake build          # ~30 min on first run; downloads + compiles Mathlib
python3 reproduce.py --check
```

Expected output: 18 proved results, zero sorry, zero admit, zero added axiom, zero native_decide calls, zero errors. The reproduce.py --check script enforces nine logical pass gates (toolchain pin, Mathlib pin, sibling artifacts, build success, zero-stub scan, theorem index, license, citation metadata, archive bundle) and verifies that the 18-theorem index in QanaryPaper6/{Basic,PositiveComposition,NegativeComposition,AdamsBridgeDiagnosis,NatEquivalence}.lean matches the listing in Appendix A.

### Artifact contents

Table 3 maps each top-level path in the artifact to its purpose.

| Path | Purpose |
| --- | --- |
| QanaryPaper6/Basic.lean | Combinatorial lemmas (filter_sub_right_eq_singleton, card_filter_sub_right_eq_one, card_filter_prod_le_mul) |

| Path | Purpose |
|---|---|
| QanaryPaper6/PositiveComposition.lean | Renewal theorem (fresh_mask_renewal), uniformity corollary, positive composition (pfpini_composition_with_fresh_mask), symmetric bound |
| QanaryPaper6/NegativeComposition.lean | Intermediate-wire multiplicity bound, contrast lemma, negative composition (composed_no_fresh_output_bound), security gap |
| QanaryPaper6/AdamsBridgeDiagnosis.lean | Five Adams Bridge instantiations (butterfly OK; Barrett leaks; fresh-mask pipeline secure; PF-PINI parameter $\max(1,2) = 2$; intermediate-uniform-with-fresh) |
| QanaryPaper6/NatEquivalence.lean | Algebraic-vs-Nat bridge (Theorem 6.1) and hardware-faithful Barrett PF-PINI(2) corollary |
| lakefile.lean, lake-manifest.json | Build configuration with pinned Mathlib and three sibling-repo dependencies (Papers 3, 4, 5) |
| lean-toolchain | Pins leanprover/lean4:v4.30.0-rc1 |
| LICENSE, CITATION.cff, .zenodo.json | MIT license, machine-readable citation metadata, Zenodo deposit metadata |
| reproduce.py | Single-command verification entry point (--check mode skips re-build) |
| README.md, DESIGN.md | Top-level documentation, design rationale, theorem index |

**Table 3.** Artifact path inventory. The five .lean source files contain the 18 theorems enumerated in Appendix A.

Paper 6 imports the Lean kernels of Papers 3 (qanaryUniversal), 4 (qanaryPaper4), and 5 (qanaryPaper5) as local-path Lake dependencies; sibling clones are therefore required for lake build to succeed (see the Reproduction block above).

**Archival DOI**

A long-term-preservation Zenodo deposit accompanies this paper, following the pattern established across the QANARY series. The **concept DOI** (always resolving to the latest archived version) is 10.5281/zenodo.19905303; the **v1.0.1 version DOI** (fixed to commit 8f17cef) is 10.5281/zenodo.19905304. Sibling-paper concept DOIs: Paper 1 10.5281/zenodo.19625392, Paper 2 10.5281/zenodo.19508454, Paper 3 10.5281/zenodo.19689480, Paper 4 10.5281/zenodo.19705450, Paper 5 10.5281/zenodo.19842166.

**Independent reproducibility**

The artifact has no external data dependencies, no random seeds, and no networked services beyond the initial lake package fetches. Verification is purely a function of the Lean 4 kernel and the pinned Mathlib commit; identical inputs yield identical outputs across machines. The trusted computing base is the Lean 4 kernel; this paper's artifact contains no native_decide invocations.

---

## Appendix A: Complete Lean 4 Proof Listing

The complete proof suite consists of 5 files, 18 public theorems, and zero sorry stubs, built on Lean 4 v4.30.0-rc1 with Mathlib at commit 322515540d7f.

| # | File | Theorem | Paper Reference |
|---|---|---|---|
| 1 | Basic.lean | filter_sub_right_eq_singleton | Lemma (§4) |
| 2 | Basic.lean | card_filter_sub_right_eq_one | Lemma (§4) |
| 3 | Basic.lean | card_filter_prod_le_mul | Lemma 4.3 |
| 4 | PositiveComposition.lean | fresh_mask_renewal | Theorem 4.1 |
| 5 | PositiveComposition.lean | fresh_mask_uniform | Corollary 4.2 |
| 6 | PositiveComposition.lean | pfpini_composition_with_fresh_mask | Theorem 4.4 |
| 7 | PositiveComposition.lean | pfpini_composition_max_bound | Corollary 4.5 |
| 8 | NegativeComposition.lean | intermediate_wire_multiplicity_bound | Theorem 5.1 |
| 9 | NegativeComposition.lean | intermediate_wire_with_fresh_is_uniform | Contrast (§5) |
| 10 | NegativeComposition.lean | composed_no_fresh_output_bound | Theorem 5.2 |
| 11 | NegativeComposition.lean | security_gap_intermediate | Theorem 5.3 |
| 12 | AdamsBridgeDiagnosis.lean | adams_bridge_butterfly_wire_ok | Theorem 7.1 |
| 13 | AdamsBridgeDiagnosis.lean | adams_bridge_barrett_wire_leaks | Theorem 7.2 |
| 14 | AdamsBridgeDiagno | adams_bridge_with_ | Theorem 7.3 |

| # | File | Theorem | Paper Reference |
|---|---|---|---|
| | sis.lean | fresh_secure | |
| 15 | AdamsBridgeDiagnosis.lean | adams_bridge_fresh_pfpini_parameter | Theorem 7.4 |
| 16 | AdamsBridgeDiagnosis.lean | adams_bridge_intermediate_uniform_with_fresh | Theorem 7.5 |
| 17 | NatEquivalence.lean | barrett_nat_equivalence | Theorem 6.1 |
| 18 | NatEquivalence.lean | barrettNat_pfpini_two | Corollary 6.2 |

The proof artifact is available at https://github.com/rayiskander2406/qanary-pf-pini-composition-arXiv-2604.25878 (tag: v1.0.1, DOI: 10.5281/zenodo.19905304, concept DOI: 10.5281/zenodo.19905303).